\documentclass[aip,pop,psfig,twocolumn,showpacs,superscriptaddress,nobibnotes,
reprint]{revtex4-1}
\usepackage{color}
\usepackage{graphics}
\usepackage{epsfig}
\usepackage[normalem]{ulem}
\usepackage{cancel}
\usepackage{float}

\usepackage[outercaption]{sidecap}

\begin{document}
\title{Interplay of single particle  and collective response  in molecular dynamics simulation of dusty plasma system}
\author{Srimanta Maity}
\email {srimanta.maity@ipr.res.in}
\affiliation{Institute for Plasma Research, HBNI, Bhat, Gandhinagar - 382428, India}
\author{Amita Das}
\affiliation{Institute for Plasma Research, HBNI, Bhat, Gandhinagar - 382428, India}
\author{Sandeep Kumar}
\affiliation{Institute for Plasma Research, HBNI, Bhat, Gandhinagar - 382428, India}
\author{Sanat Kumar Tiwari}
\affiliation{Indian Institute of Technology Jammu, Jammu, Jammu and Kashmir, 181121, India}
\begin{abstract} 
\paragraph*{}

Collective response of the plasma medium is well known and has been explored extensively in the context of dusty plasma medium. On the other hand,  individual  
particle response associated with the  collisional character  giving rise to dissipative phenomena  has not been explored adequately. 
In this paper two-dimensional (2-D) molecular dynamics (MD) simulation of 
dust particles interacting via  Yukawa potential  has been considered. It has been shown that disturbances induced in a dust  crystal 
 elicit both collective and single particle responses. Generation of a few  
   particles moving at  speeds considerably   higher than acoustic and/or  shock speed ( excited by the  
   external disturbance) are observed.  This is an indication of a single particle response.  Furthermore, as these individual energetic particles propagate,  
   the dust crystal is observed to crack  along their path. Initially when the energy is high  these particles    generate secondary energetic particles by 
   collisional scattering process.   
   However, ultimately as these particles slow down they  excite collective response in the dust medium at 
   secondary locations in a region which  is undisturbed by the primary external disturbance.  
   The condition  when the cracking of the crystal stops and collective excitations get initiated 
    has been  identified quantitatively.  The   trailing collective primary disturbances would thus often encounter a disturbed medium with secondary and tertiary 
    collective perturbations, thereby suffering significant modification in its propagation.  
    It is thus clear that there is an interesting interplay (other than mere dissipation) 
    between the single particle and collective response  which governs the dynamics of any disturbance 
    introduced in the medium.

\end{abstract} 
\maketitle 
\section{Introduction}
\label{intro}

Dusty plasma is a multicomponent plasma which contains electrons, ions, neutrals and dust grains \cite{shukla2015introduction} (of nano \cite{deka2017observation, praburam1996experimental, takahashi1998analyses, winter1998dust, greiner2012imaging, thomas2012magnetized} to micrometer scales \cite{barkan1994charging, quinn1996structural, klindworth2000laser, fortov2003dynamics}). Electrons with higher mobility compared to ions,  bombard and stick to the dust surface, causing the dust grains to acquire high negative charges. These charged dust particles in the plasma environment act as a third species. 
A  micron-size dust grain typically acquires a  charge of the order of $10^4-10^5$ elementary charges. \
The high value of dust to ion mass ratio ($m_d/m_i$), as well as a very low charge to mass ratio of the dust $Q_d/m_d$ compared to the other two species, makes the response time of the  dust species to be  much longer  compared to that of  electrons and ions. 
Thus, while tracking the dust evolution the response of electron and ion species can be considered as an instantaneous inertia-less response. 
The number density of electron and ion species can then be taken to follow the Boltzmann distribution. 
The charge on individual dust grains is assumed to get shielded instantaneously by the lighter electron and ion species of the plasma. 
The inter-dust interaction potential is thus a  
shielded Coulomb  potential \cite{shukla2015introduction} of the form $({Q_d^2}/{4\pi\epsilon_0r})\exp(- r/\lambda_D)$ which is also termed as Yukawa 
potential. Here  $Q_d$ defines the charge on the dust and   $\lambda_D$ is the plasma Debye length.

The presence of charged dust grains as an extra component,  introduces a rich variety of low-frequency collective modes \citep{berkovsky1992spectrum, kaw1998low} in the plasma. 
The characteristic eigenmode frequencies of the medium being low the dust dynamics can be easily visualized even by a naked eye. 
The ratio of the average inter-grain potential energy to the dust thermal energy defines the coupling parameter, $\Gamma = {Q_d^2}/{4\pi\epsilon_0a k_B T_d}$, where $k_B$ is the Boltzmann constant,  $T_d$ represents the temperature of the dust and $a$ represents inter-particle distance. We define $\Gamma_{eff} = \Gamma\exp(-\kappa)$ by including the shielding factor in the interaction potential in our Yukawa model, where $\kappa = a/\lambda_D$ is the screening parameter.
The high charge $Q_d$ on the dust ensures
that the dust medium can be easily found in a strongly coupled state \cite{murillo2004strongly, chu1994direct} and requires no stringent criteria on temperature $T_d$ and density (through the dependence on $a$) to be satisfied. 
 The dusty plasma, thus,  exhibits a wide variety of phases from gaseous to liquid and to crystalline state. It has also been found to behave like complex visco - elastic fluids 
 \cite{kaw1998low}. 
 The unique property ( response at human perceived time and length scales )  of the dusty plasma along with the ease with which it can be 
  prepared in a strongly  coupled  regime, 
 makes it an ideal medium to explore dynamical processes in different phases of matter including complex fluids. Particle level dynamics leading to macroscopic correlated phenomena can be directly observed and analyzed, for example, the phenomena of phase transitions  \cite{phasetransitions}, diffusion, and viscous effects  etc., can be tracked as it is happening.  \

The collective dynamical properties displayed by the dusty plasma medium have been investigated thoroughly by many authors.  These include linear excitations ( e.g. dust acoustic waves \cite{rao1990dust, shukla2001survey}, dust lattice wave \cite{melandso1996lattice, pieper1996dispersion, misawa2001experimental}, dust ion acoustic waves \cite{Ion-acoustic, barkan1996experiments} and so on), nonlinear response (dust acoustic solitons \cite{sharma2014head, jaiswal2014theoretical, harvey2010soliton, boruah2015oblique, bandyopadhyay2008experimental}, shocks \cite{shukla2001dust, shukla2003solitons, PhysRevE.64.066407, nakamura1999observation}, Korteweg - de Vries (KdV) solitons \cite{kumar2017observation, shukla2003solitons, nakamura2001observation}, instabilities \cite{tiwari2012kelvin, das2014suppression, samsonov1999instabilities, merlino1998laboratory, rosenberg1993ion}).   The wakes created behind a moving disturbance \cite{nosenko2002observation, miloch2010wake, sundar2017impact, ludwig2012wake, block2012wake, ludwig2014introduction}, Mach cones \cite{Machconeshocks, samsonov1999mach, hutchinson2003ion}, precursor solitons or shocks \cite{Forewake, jaiswal2016experimental} in flowing dusty plasma, have also been reported. On the other hand, the scattering phenomena where particle effects dominate instead of collective features, has been studied only recently by Murillo et al. \cite{marciante2017thermodynamic}, for the  Yukawa system.
 
 In this work, we illustrate the interplay between single particle effects and collective phenomena in the context of dusty plasma depicted by Yukawa interaction through MD simulations.  
This paper has been organized as follows. In Section \ref{mdsim} we have described the  MD Simulations in detail. In Section \ref{reslt} we present our observations which show cases when single particle effects play a crucial role in describing the course of collective phenomena.  
 Finally, Section \ref{Summry} provides the summary of our work.
 
 \paragraph*{•}

\section{Simulation details}
\label{mdsim}
An open source MD code, Large-scale Atomic/Molecular Massively Parallel Simulator (LAMMPS) \cite{plimpton1995fast} has been used for present simulations. 
 The dust particles  interact with Yukawa (screened Coulomb) potential. The contribution of lighter electron and ion species 
 is taken into account in the screening factor. We carry simulations in a  2-D (X-Y) plane. The dust density is related to the Wigner Seitz radius in 2-D by the relationship $ a=(\pi n_{2d})^{(-{1}/{2})} $. The typical  parameters \cite{nosenko2004shear} 
 chosen for our simulation studies are as follows: (i)  the mass of the dust grain  $m_d = 6.99\times10^{-13} $ Kg, (ii) charge on dust $Q_d = 11940e$ (where $e$ is an electronic charge) and (iii) $a = 5.6407\times 10^{-4}$ m. The typical density corresponding to  this choice of $a$, the average inter-particle 
 separation is $n_{d0} = 1.0\times 10^6$ $m^{-2}$. We have chosen the Debye length 
  $\lambda_D  = 5.6407 \times10^{-4}$ m which is same as $a$. 
  A  two-dimensional periodic system of $20000$ charged point particles are placed in a simulation box length of   $L_x = 177a$ and $L_y=354a$ along $X$ and $Y$ directions, respectively. In our simulation, the cut-off distance for the inter-grain interaction is kept to be at $20a$. For these parameters the characteristic dust plasma frequency, $\omega_{pd} = \sqrt{{Q_d^2}/{2 \pi \epsilon_0 m_d a^3}} \simeq 22.8630$ $s^{-1} $, corresponds to the dust plasma period $0.2749$ s. We have chosen our simulation time step to be of $0.01\omega_{pd}^{-1}$, which ensures that the phenomena associated with the dust dynamics can be easily resolved. 
   From henceforth wherever it is not explicitly mentioned the time and length scales would be assumed to be normalized by $\omega_{pd}^{-1}$ and  $a$ 
   respectively. \\

Initially, the equilibration of this 2-D dust grain system has been achieved using a Nose-Hoover \cite{nose1984molecular, hoover1985canonical} thermostat by distributing the particles in canonical (NVT) ensemble. The assigned equilibrium temperature has been obtained by running NVT thermostat for $1000\omega_{pd}^{-1}$. After that, we have disconnected canonical thermostat and let the system evolve in the presence of a microcanonical (NVE) thermostat for an additional  time of
$1000\omega_{pd}^{-1}$. This ascertains that the system finally reaches a  thermodynamic equilibrium at the desired temperature. For all our studies, we have chosen $\Gamma_{eff}$ and $\kappa$ as  $250 $ and $1.0$ respectively.


\section{Observations}
\label{reslt}
\begin{figure}[hbt!] 
\includegraphics[height = 4.0cm,width = 9.0cm]{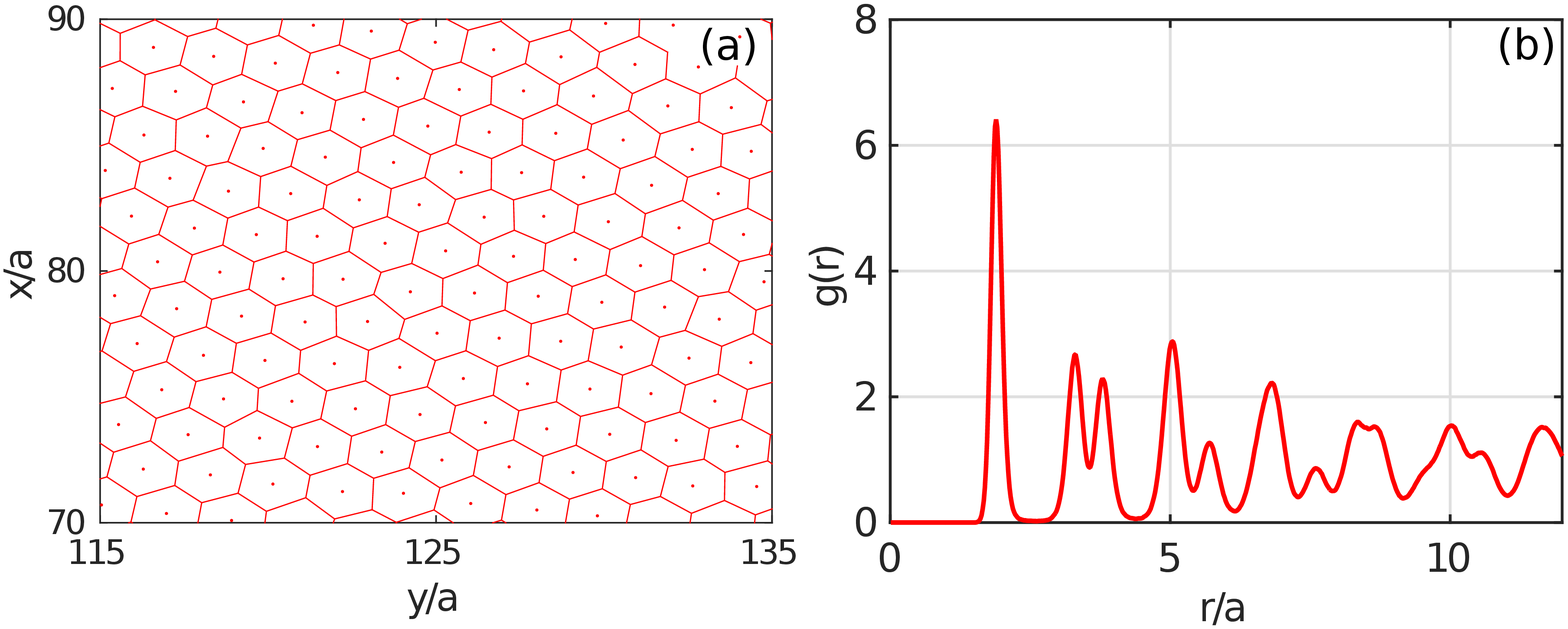}
                   \caption{(a) Voronoi tesselation clearly depicting the formation of a nearly hexagonal crystalline structure in equilibrium and (b) Sharp multiple peaks in radial distribution function (rdf) of the system of particles (dust grains) in equilibrium also confirm the crystalline state of the medium. For both the plots, we have taken $\Gamma_{eff} = 250$ and  $\kappa = 1.0$.}
                  \label{equilibrium}    

\end{figure}

\begin{figure}[hbt!] 
\includegraphics[height = 12.0cm,width = 8.5cm]{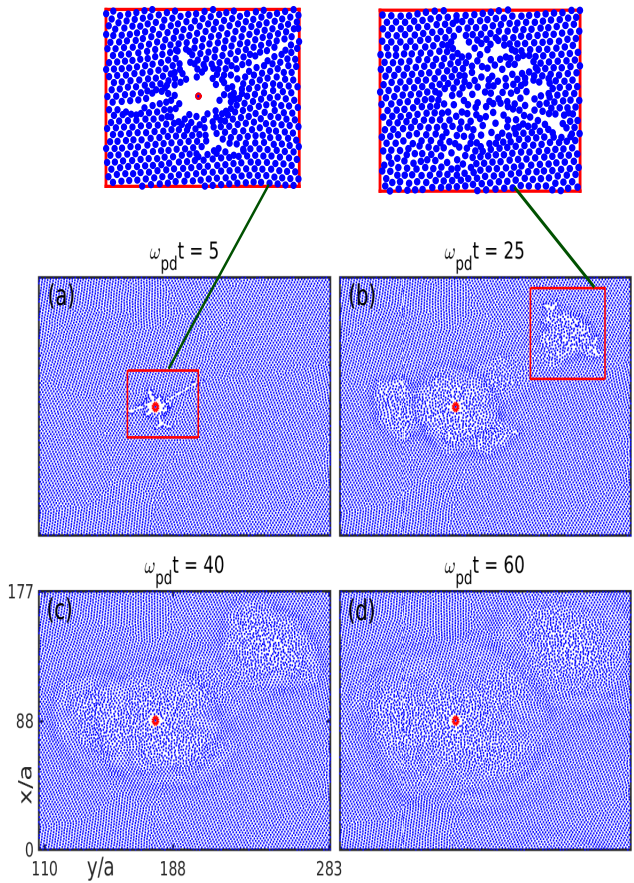}
                   \caption{Time lapse sequence of particle configuration in position space after adding an extra particle with charge, $Q_p = 100Q_d$ in the system. Generation of cracks and deformations in the crystal structure made by some high energetic particles traveling through the crystal is shown in (a). The excitation and propagation of collective modes are shown in (b)-(d).}
                  \label{position space diagram}    

\end{figure}

\begin{figure}[hbt!] 
\includegraphics[height = 12.0cm,width = 9.0cm]{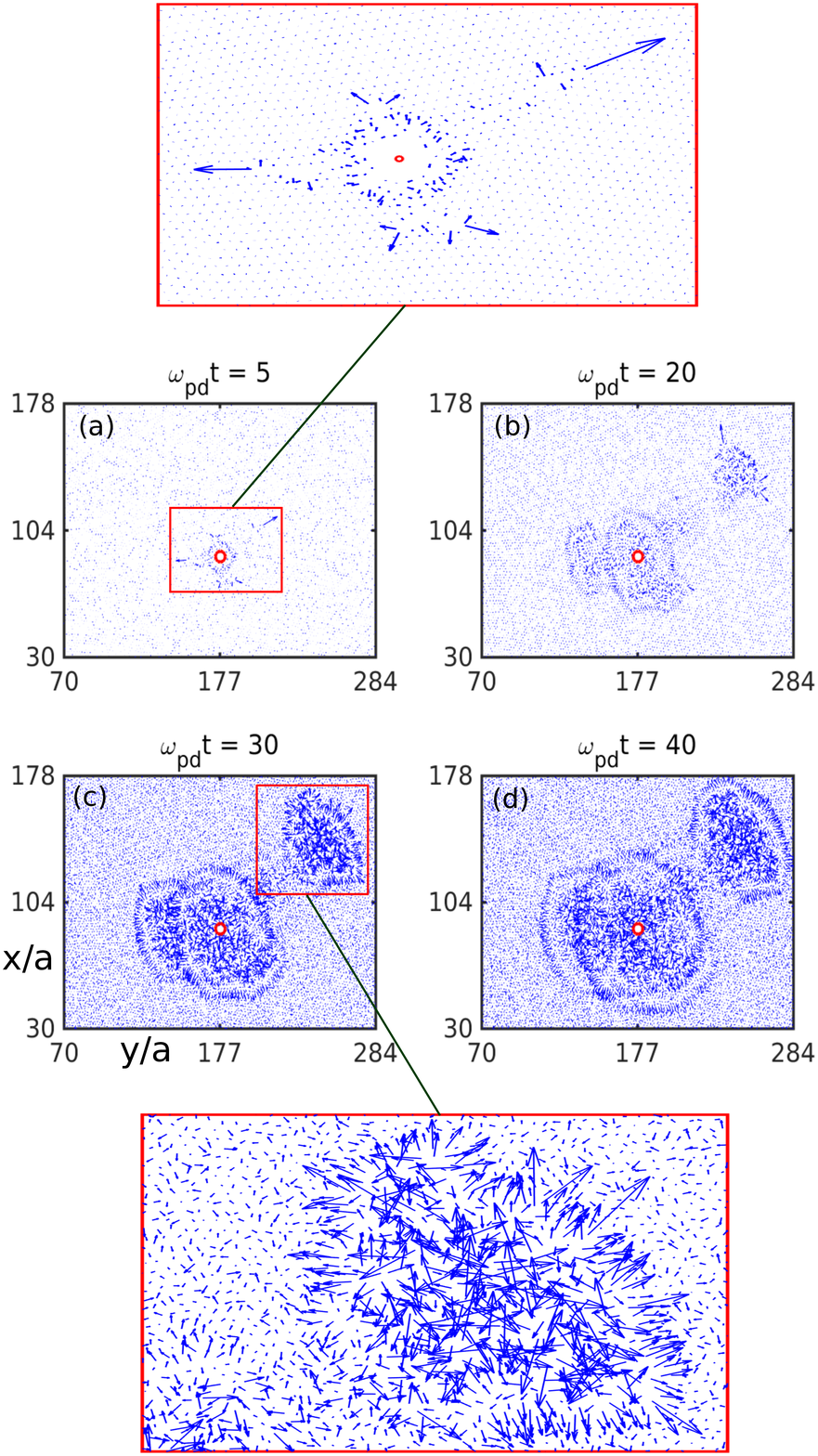}
                   \caption{Time evolution of the velocity of particles. Length of the arrow represents velocity amplitude of the particle. The ejection of high energetic particles in random directions is shown in (a). These high energetic particles travel through the medium creating cracks and defects in the crystal and finally generates  collective disturbances far from the initial perturbation as shown in (b)-(d).}
                  \label{velocity_quiver_single}    

\end{figure}

A nearly hexagonal Yukawa crystal (shown in Voronoi diagram in Fig. \ref{equilibrium}(a)) has been obtained for $\Gamma_{eff} = 250$ and $\kappa = 1.0$ using MD simulations. Multiple peaks in radial distribution function (rdf) (Fig. \ref{equilibrium}(b)) indicate the crystalline phase of the medium in the equilibrium. \
We now add an extra particle to  the medium with 
 a charge of $Q_p = f Q_d$ (i.e. the charge of the inserted particle is chosen to be $f$ times the charge of the individual dust grains.  In our simulations, $f$  has been varied from 
 a value of unity to $100$). The addition of this particle can be considered as similar to inserting a  probe with a biased voltage in the medium experimentally.  The insertion of this particle disturbs the equilibrium and triggers a response from the dust medium.  It is 
  observed that when the charge of this external particle 
  is much higher compared to the background dust particles, a few dust particles 
 suffer rapid displacement (Fig. \ref{position space diagram}(a)). 
  The collective response of the medium is also observed to be present (Fig. \ref{position space diagram}((b)-(d))), however,  the speed of the particles involved in this is comparatively very slow. The velocity quiver plot of Fig. \ref{velocity_quiver_single}((a)-(d))  also clearly shows that a
   few particles move very rapidly.  Their movement through the medium generate cracks \cite{hutchinson1991mixed} and defects in the crystal structure along their path. Subsequently, they seem to loose their kinetic energy and significantly slow down. At their slow phase,  
  these particles form secondary centers from where the collective response of the dust medium emanates. These secondary centers are at locations which are significantly separated from the region where the external particle was introduced.  
  This is clearly evident from Fig. \ref{velocity_quiver_single}((b)-(d)).
The highly energetic particles, which are few in number,  get generated as soon as the medium is disturbed. They are essentially single particle scattering response from the potential disturbance introduced by inserting the extra particle.  

\begin{figure*}[hbt!] 
\includegraphics[height = 18.0cm,width = 18.0cm]{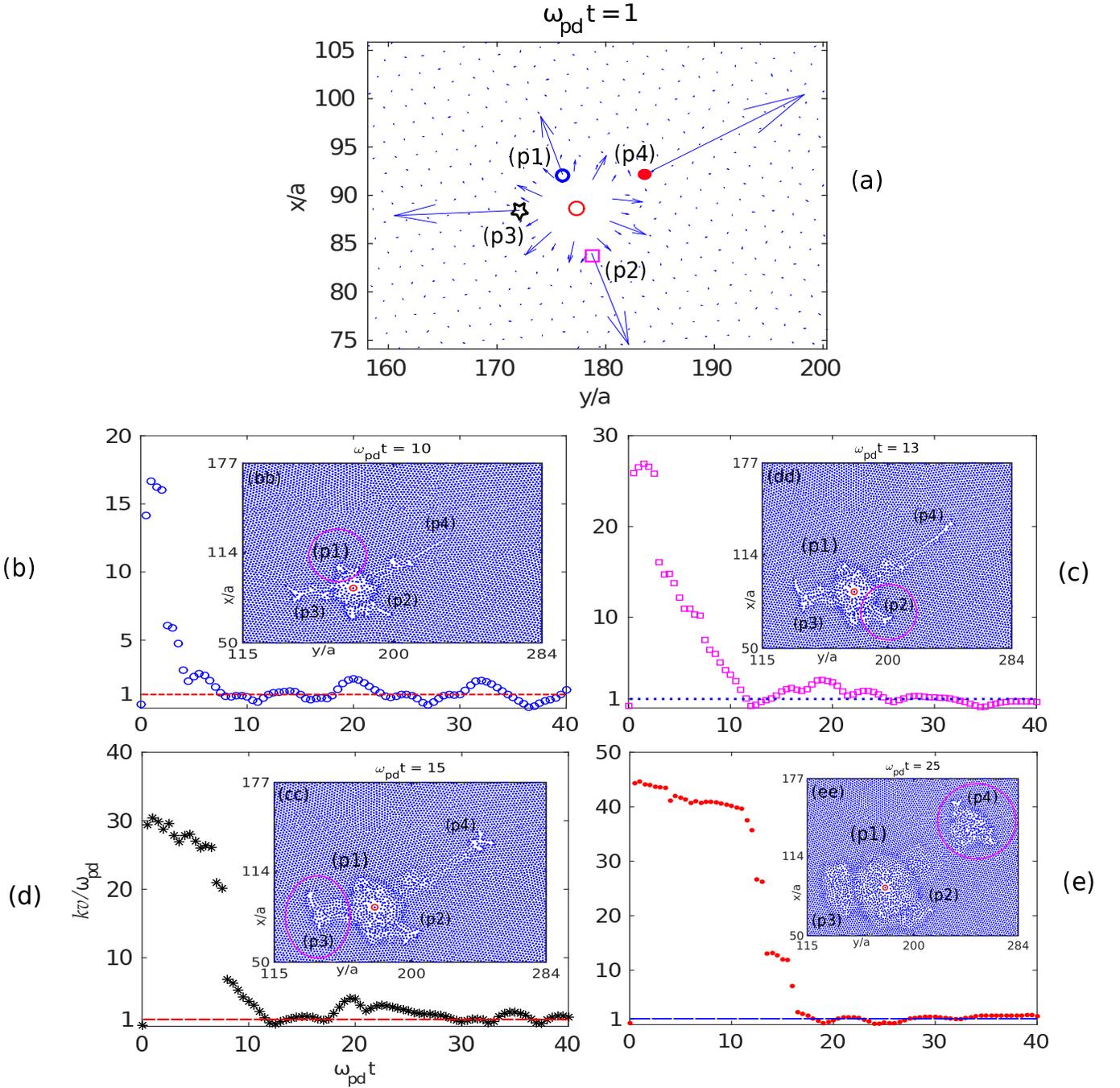}
                   \caption{(a) Shows the velocity quiver plot of particles at $\omega_{pd}t = 1$, after inserting an extra charged particle (red marked circle) with $Q_p = 100Q_d$ in the medium. Here the length of the arrow represents the relative amplitude of the velocity of particle. (b)-(e) show the time evolution of frequencies associated with perturbations made by high energetic particle, ($p1$), ($p2$), ($p3$) and ($p4$) respectively. Insets (bb)-(ee) represent the particle configurations, showing the excitation of collective disturbance (pink marked region) caused by each of these high energetic particles ($(p1)$, $(p2)$, $(p3)$ and $(p4)$), respectively.  }
                  \label{single_particle}    
\end{figure*}
 The velocity quiver plots (Fig. \ref{velocity_quiver_single}) show that  few particles are ejected in random directions with very high velocities. The fastest particle 
  reaches furthermost from the original disturbance as can be observed from zoomed plot of 
 Fig. \ref{velocity_quiver_single}(a)  shown at $\omega_{pd}t = 5$.  
 As these energetic particles trace their way through the crystal, they interact with the lattice. During the initial phase, they seem to generate cracks in the crystal and also tend to generate secondary energetic particles. 
 As a result of such encounters, the original particle subsequently slows down. It is observed that after losing a significant amount of energy it creates a secondary center of collective excitation in the medium (Fig. \ref{velocity_quiver_single}(c)).

 We now try to understand the condition when  the ballistic propagation of the energetic particle stops and  it excites 
   collective disturbance in the medium.  For this purpose, we chose to track four distinct particles in the order of increasing energy 
( created by the initial external disturbance in the medium). They are marked as $p1, p2, p3, p4$ and are identified by in Fig. \ref{single_particle}(a) by various symbols. The size of the arrow associated with these four particles has also been drawn which indicates their respective speeds $v$. 
 In Fig. \ref{single_particle}((b)-(e)) we show the evolution of $ R=kv/\omega_{pd}$ for all the four particles. Here  $k = 2\pi/a$ is the wavenumber associated with the lattice spacing  $a$ 
 and $v$ is the speed of the particle that one is tracking. 
Hence,  $(kv)^{-1}$  is the typical time scale associated with the energetic particles.  When the collective modes of the system have similar response time, then they can get excited. 
  It is observed that initially the value of $R=kv/\omega_{pd}$ is much higher than unity. After that there is a steady decrease of $kv/\omega_{pd}$ 
for each particle and ultimately it reaches the value of unity for each of them and hovers around it. Thus only when the time scale associated with the energetic particle movement is similar to the collective response of the medium, the collective modes get excited. 
The inset of the subplots ((b), (c), (d), (e)) of Fig. \ref{single_particle} demonstrates it in a clear fashion. 
In subplot (b) of Fig. \ref{single_particle} we have shown the evolution of  $R_{p1}$, where $p1$ is the particle with slowest speed. It can be observed that $R_{p1}$ 
has already reached the value of unity before  $ \omega_{pd}t \sim 10 $.  For other three particles (viz., $p2, p3, p4$) $R_{pi} > 1$ for ($i = 2,3,4$). The inset of Fig. \ref{single_particle}(b) shows that while the collective response at the location of $p1$ at $\omega_{pd}t = 10$ has 
already been initiated , the other three particles are still marching ballistically ahead in the crystal. This gets further confirmed from the inset of other subplots ((c), (d), (e)).  
For instance,  in Fig. \ref{single_particle}((c), (d)) the inset shows the particle picture at $\omega_{pd}t = 13$ and $15$. At these times $R$ for all the three particles are close to unity except the $4$th particle. It should be noted that the crystal shows collective response around all the three particles in these subplots, except the fourth one. 
The $4$th particle is still moving ballistically ahead. In subplot (d) the inset is shown at $\omega_{pd}t= 25$. 
At this time even  $R_{p4}$ has touched the value of unity and it can be observed that there are secondary collective disturbances around this $4$th particle as well. 
 The secondary center excited by each of the energetic particle forms at a random location depending on the angle and velocity with which they got scattered initially.

\begin{figure}[hbt!]  
 \includegraphics[height = 7.0cm,width = 8.0cm]{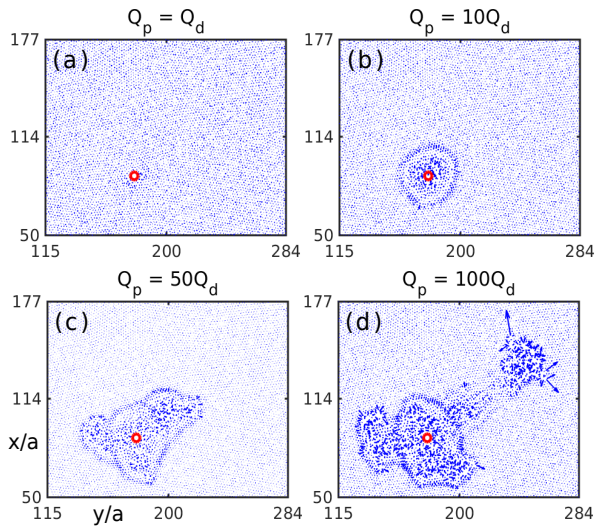}
                   \caption{velocity quiver diagrams of particles after inserting an extra particle in the medium with charge (a) $Q_p = Q_d$, (b) $Q_p =10Q_d$, (c) $Q_p = 50Q_d$ and (d) $Q_p = 100Q_d$ at $\omega_{pd}t = 20$. Almost no disturbance is made in the system as we insert the extra particle having same charge as that in the medium as shown in (a). Collective disturbance propagating isotropically around the point of intertion as shown in (b). Disturbances get anisotropic for higher amount of charges as shown in (c) and (d).}
                  \label{velocity_Quiver_varying_charge}                            
\end{figure}

We have also investigated the role of the initial strength of perturbation on both the collective and single particle features by varying the charge of the extra particle which was added in the medium. It can be clearly seen from the velocity quiver plots shown at $\omega_{pd}t = 20$ in Fig. \ref{velocity_Quiver_varying_charge} 
that when we put the extra particle having a charge equal to the dust grains ($Q_p = Q_d$, 
i.e. $f=1$) there is hardly any disturbance which persists in the medium (Fig. \ref{velocity_Quiver_varying_charge}(a)).  This is expected,  as the extra particle is same as the background dust grains of the medium and
hence they easily adjust in the equilibrated system. When the charge is ten times the charge of the background dust,  merely collective response around the extra charge is observed to develop. The symmetric form of the collective disturbance around the added particle (Fig. \ref{velocity_Quiver_varying_charge}(b)) bears testimony to this.

However, with increasing charge (e.g. $f = 50, 100$)   the disturbances get  anisotropic.  The shape is essentially governed by the paths taken by  
 the few energetic particles which get generated by the individual scattering events, and which ultimately trigger the collective response from  secondary centers when 
 their speeds get slower significantly to match with the time scale of collective response. A comparison 
of $f = 50$ and $f = 100$ shows that higher the charge of the added particle,  the secondary center 
forms at a location which is further away from the originally inserted particle. This is essential because the maximum energy acquired by the individual scattering events increases with the charge of the particle that has been added in the medium. 



\begin{figure}[hbt!]  
 \includegraphics[height = 9.0cm,width = 9.0cm]{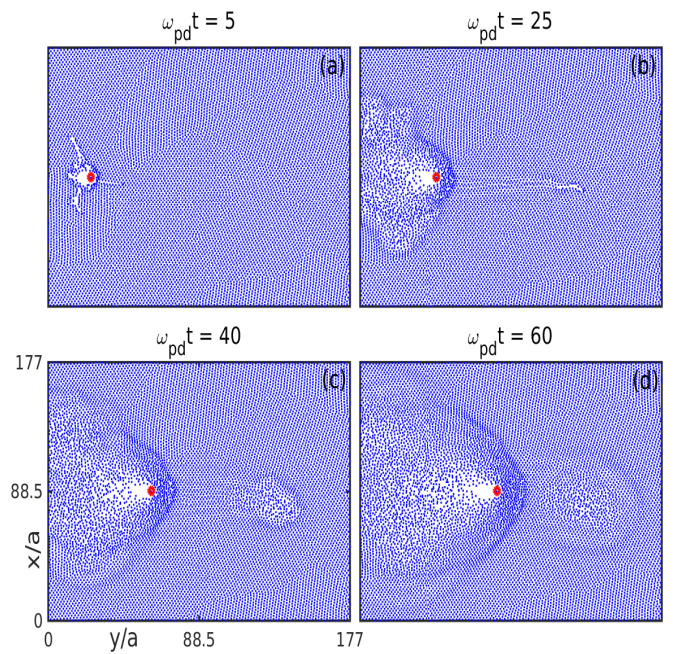}
                   \caption{Time lapse sequence of the snapshots of particle configurations when a projectile (red marked circle) with charge $Q_p = 100Q_d$ and velocity $V_{yp} = 1.3\times10^{-2}$ m/sec (Mach number, $M =1.41$ ) moving through the medium. Single particle scattering is shown in (a) and (b) -(d) shows the occurrence of deformations and disturbances far ahead of the precursor shock.}
                  \label{position_space_projectile}                            
\end{figure}
We believe that these observations, in turn, would have far reaching consequences to many kinds of collective phenomena that has so far been observed in the context of dusty plasma medium. For instance, it is well known that for a 
  plasma which is flowing around an obstacle, or alternatively when a 
highly charged projectile moves through a plasma, shocks are formed ahead of the projectile if the projectile velocity exceeds the acoustic speed. 
These shocks in general move with the speed of the projectile.    A charged projectile in the medium should also elicit single particle scattering events leading to energetic particle generation.  These faster particles can disturb the medium ahead of the shock region which otherwise should have remained undisturbed. We now carry out simulations to ascertain whether this indeed happens. 
%

\begin{figure}[hbt!] 
  \includegraphics[height = 8.0cm,width = 8.5cm]{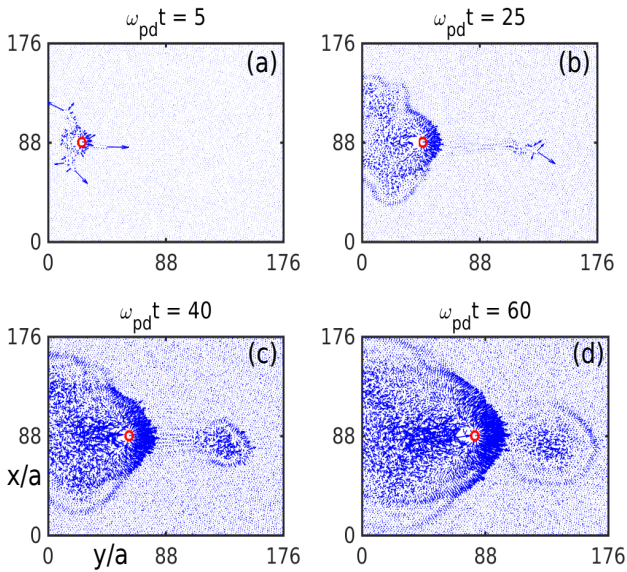}
                  \caption{ Time series of velocity quiver plots with a projectile (red marked circle) having charge $Q_p = 100Q_d$ and velocity $V_{yp} = 1.3\times10^{-2}$ m/sec ($M = 1.41$). }
                  \label{Velocity_quiver_projectile}
  \end{figure}

We insert a charged particle in the system with $Q_p = 100Q_d$ and a velocity $V_{yp} = 1.3\times10^{-2}$ m/sec, associated with the Mach number, $M = 1.41$. The Mach number is defined by the ratio of the projectile velocity to the dust acoustic speed of the medium, i.e., $M = V_{yp}/C_s$. The dust acoustic speed is first calculated as follows: At first, a linear (small amplitude) electric field perturbation (along $\hat{y}$) is given to the system to excite dust acoustic wave. Then from the slope, ($dy/dt$) of the plot of the trajectory along $\hat{y}$ (after averaging in $x$) with respect to time, we have calculated the acoustic speed of the medium. It turns out that the dust acoustic wave speed ($C_s$) of the medium for $\Gamma_{eff} = 250$ and $\kappa = 1.0$ is equal to the $9.2\times10^{-3}$ (m/sec).
 It can be observed from Fig. \ref{position_space_projectile}, that at $ \omega_{pd}t = 5$ the dust particles get evacuated from the neighborhood of the projectile and a few dust grains acquire high velocities.  They move very rapidly away from the projectile. This is more apparent from the velocity quiver plot of Fig. \ref{Velocity_quiver_projectile}. 
A shock structure is also observed to form ahead of the projectile. However, the energetic particles disturb the unshocked crystalline medium beforehand. Thus when the shock region catches up, it encounters not the original medium but a disturbed crystal.  
In these simulations also,  energetic particle has a  ballistic propagation in the medium in the beginning wherein it creates cracks and defects along its path. 
Subsequently, however, as it slows down it excites a collective response in the medium around a   point which is located considerably apart from the position of the projectile. 
  \begin{figure}[hbt!] 
\includegraphics[height = 10.0cm,width = 8.0cm]{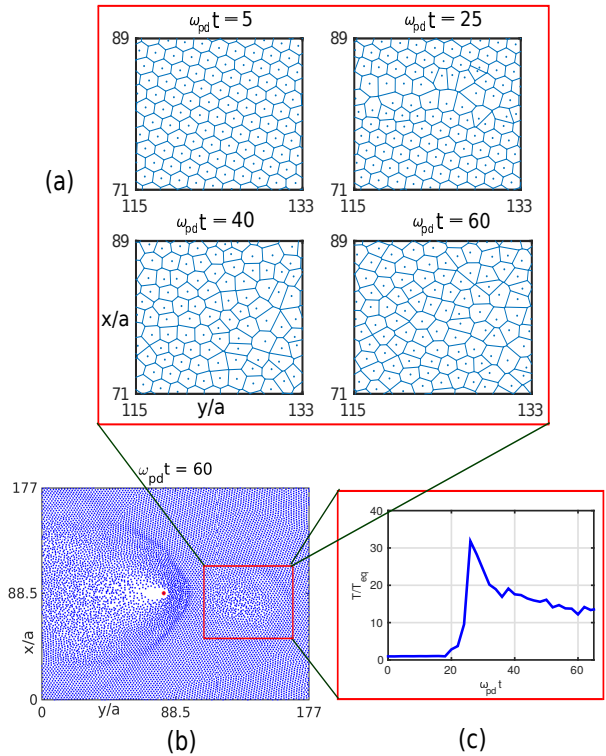}
                   \caption{Occurrence of phase transitions in a local region (red marked quadrilateral in (b)) far-ahead of the precursor. Time series of Voronoi diagram indicates the breaking of crystal symmetry, as shown in (a). Time evolution of temperature of the local region is shown in (c). }
                   
                   \label{diagram_temp}
                  
\end{figure}

The collective excitations at these secondary centers change the properties of the medium in the neighborhood. We have illustrated this with the help of Voronoi plots (Fig. \ref{diagram_temp}(a)). The initial lattice has an ordered hexagonal form. However, after it gets disturbed, the lattice structure breaks down and system appears to take a disordered form. 
 We have also evaluated the temperature of the localized region disturbed by one of the energetic particles much ahead of the projectile.  
 In Fig. \ref{diagram_temp}(c)  the time evolution of the temperature of this localized region has been shown.  It is observed that the  temperature of this region  increases almost $30$ 
 times when the energetic particles reach there.  Subsequently,  however, the temperature appears to steadily decrease. However, it still remains quite high (about 10 times) when the shock associated with the projectile arrives at this location. This would correspond to an effective $\Gamma_{eff} = 25$ for this local region. 
 This value of $\Gamma_{eff}$ corresponds to an intermediate phase of complex fluid between liquid and solid states, which has often been characterized as visco - elastic medium. 
 Thus, instead of a regular crystal structure, the shock propagating with the projectile would encounter a region of visco-elastic medium rather than a crystal state. Thus, the propagation of precursor solitons and/or dispersive shock, which are moving ahead of the projectile at comparatively slower time scale,  will encounter patches of disturbed medium in an altogether different phase. This will effect their dynamics considerably. 
 
 \begin{figure}[hbt!] 
 \includegraphics[height = 8.cm,width = 9.0cm]{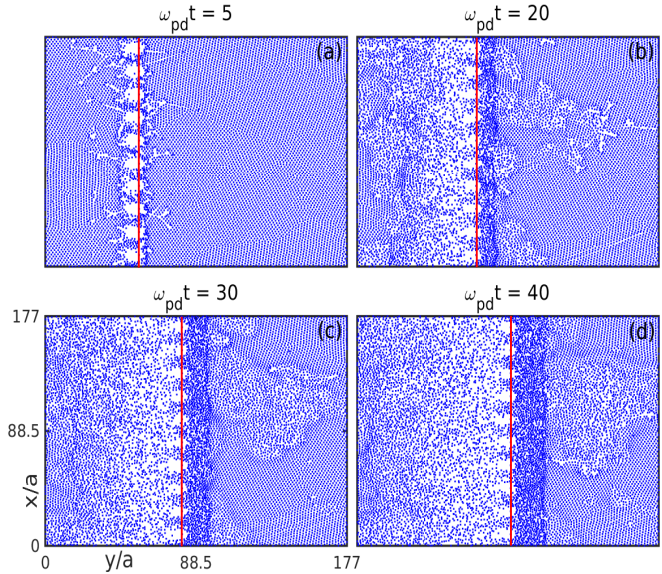}
                 \caption{Time series of particle configurations of the system when an array of 10 particles (red marked line) each with charge, $Q_p = 100Q_d$ and velocity, $V_{yp} = 1.3\times10^{-2}$ m/sec (associated Mach number, M = 1.41) moving along $\hat y$ direction through the medium.}
                 \label{position space of wire moving}
\end{figure}
 
 We have also simulated a case with multiple projectiles, all of them  \cite{Forewake} having the same high charge ($Q_p = 100Q_d$) and moving with the same velocity along $\hat{y}$. This choice helps preserve periodicity condition along $\hat x$ direction. Furthermore, with an increased number of such projectile particles, one essentially mocks up a  moving wire having a   potential bias (frequently used in experiments \cite{jaiswal2016experimental}). For this case too one observes a few energetic particle shoot ahead of the projectile initially creating disturbances in the medium ahead. A planar shock gets formed which then encounters a disturbed medium. This has been shown in Fig. \ref{position space of wire moving}.
 
\section{Summary}
\label{Summry}
 We have considered a dusty plasma medium for which the 
 dust grains interact amongst themselves via Yukawa interaction. 
  We have carried out MD simulations to investigate the response of such a  dusty plasma medium to an imposed disturbance. We 
created disturbance in the medium  by adding an extra charged (unity to $100$
times the charge in each dust particles) 
particle (static as well as moving in the medium with a specified velocity). It is observed that medium has two distinct modes of response. The initial fast response arises through individual particles scattering by the imposed potential. Subsequently,  the medium also shows the collective response in terms of acoustic waves, 
foreshock generation etc. It has been observed that the 
scattering response can often lead to a generation of a few very energetic particles. These particles move very rapidly in the medium. Such fast particles, as they move ballistically in the medium introduce cracks and defects in their paths. However, when they slow down they invoke the collective response in the medium in localized regions. These locations have been termed as the secondary centers by us. 
The primary centers of collective response triggered directly by the additional high charged particles subsequently catches up with these secondary centers and encounters a disturbed medium. This influences their subsequent development. 

It has thus been shown by our simulations that the dusty plasma responds to any externally imposed disturbances in two distinct ways. These are single particle 
scattering events 
(often leading to the generation of energetic particles) as well as the collective response. Subsequently, however, the overall evolution strongly depends on the interplay of these two responses in the medium. \\


%



\bibliography{cracking_ref}
\end{document}